\documentclass[a4paper]{article}

\usepackage{INTERSPEECH2020}
\usepackage{dirtytalk}
\usepackage{amsmath}
\usepackage{caption}
\usepackage{subcaption,color}
\usepackage{enumitem}
\usepackage{graphicx}
\usepackage{dblfloatfix}
\DeclareMathOperator*{\argmax}{arg\,max}

\title{Improving Tail Performance of a Deliberation E2E ASR Model Using a Large Text Corpus}
\name{Cal Peyser, Sepand Mavandadi, Tara N. Sainath, James Apfel, Ruoming Pang, Shankar Kumar}
\address{Google Inc.}
\email{\{cpeyser,sepand,tsainath,japfel,rpang,shankarkumar\}@google.com}

\begin{document}

\maketitle
\begin{abstract}
End-to-end (E2E) automatic speech recognition (ASR) systems lack the distinct language model (LM) component that characterizes traditional speech systems.  While this simplifies the model architecture, it complicates the task of incorporating text-only data into training, which is important to the recognition of tail words that do not occur often in audio-text pairs.  While shallow fusion has been proposed as a method for incorporating a pre-trained LM into an E2E model at inference time, it has not yet been explored for very large text corpora, and it has been shown to be very sensitive to hyperparameter settings in the beam search.  In this work, we apply shallow fusion to incorporate a very large text corpus into a state-of-the-art E2E ASR model.  We explore the impact of model size and show that intelligent pruning of the training set can be more effective than increasing the parameter count.  Additionally, we show that incorporating the LM in minimum word error rate (MWER) fine tuning makes shallow fusion far less dependent on optimal hyperparameter settings, reducing the difficulty of that tuning problem. 
\end{abstract}
\section{Introduction}
\label{sec:intro}

Rare words pose an ongoing problem to building high-quality speech recognition systems.  Since rare words are likely to be named entities such as names and locations, these \say{tail} words are often critical to the meaning of the decoded transcript.  Since they do not occur often in the audio-text pairs that comprise an ASR system's training set, they are difficult to predict correctly.

Conventional ASR systems contain separate acoustic, pronunciation, and language models which are run one after another.  In such systems, the distinct language model provides an opportunity to train part of the model on text-only data, which is often far more plentiful than audio-text pairs, and can contain many occurrences of words that are rare in the acoustic data.  The independence of the LM from the ASR system allows its dataset or training procedure to be adapted to specific domains, including tail words \cite{Sak13, Huang10}.

E2E ASR systems consist of a single neural network in which all components are jointly trained.  These models offer the advantage of simplifying the alignment of audio to text \cite{Wang19}, as well as decreased model size \cite{He18}. However, there is no explicit LM in an E2E architecture, complicating the task of integrating text-only data.  Many \say{LM fusion} methods have been proposed, including \say{shallow fusion} \cite{Caglar15}, in which LM logits are interpolated with those of an E2E model during inference, as well as more sophisticated methods such as \say{deep} and \say{cold} fusion, in which the LM is incorporated into the neural architecture of the E2E system \cite{Caglar15, Sriram17}.  In \cite{Toshniwal18}, shallow fusion was shown to be the most effective fusion method with a state-of-the-art E2E system, although the \say{density ratio method}, has been shown to outperform shallow fusion for a domain transfer scenario~\cite{McDermott19}.

Earlier works on shallow fusion such as \cite{Chorowski16} and \cite{Battenberg17} use LMs taken from Kaldi \cite{Povey11} recipes which are trained on no more than a few hundred million words.  Of course, language models can scale to far larger datasets.  \cite{Rafal16} trained an RNN-LM on the One Billion Word Benchmark \cite{Chelba13}, while  \cite{Radford19} trained a transformer on 8 million web documents totaling 40GB of text.  To our knowledge, the study that uses the most text data in training an RNN-LM for shallow fusion to date is \cite{Kannan18}, which uses about 4 billion words.  

The research has also shown that shallow fusion is difficult to implement correctly.  In \cite{Chorowski16}, it is shown that without careful tuning of several hyperparameters, shallow fusion causes transcripts to be cut off after only a few words, massively degrading performance. 

We have two goals in this work.  First, we seek to reduce the difficulty of tuning fusion hyperparameters.  We show that applying shallow fusion during minimum word error rate (MWER) training adapts the model to a particular setting of hyperparameters, and almost eliminates the impact of those parameters in inference.  Second, we seek to scale shallow fusion to a text corpus of about 50 billion words, an order of magnitude larger than \cite{Kannan18}.  We show that tail performance can be improved by careful pruning of the dataset without resorting to extremely large model sizes.

In this study, we focus on the particularly difficult problem of tail words.  We use shallow fusion to incorporate an LM into an E2E model trained in the recent deliberation framework \cite{Hu20}, which already achieves state-of-the-art transcription quality on rare words.  We show that shallow fusion with a large text corpus yields further improvements on the tail.

The rest of this paper is organized as follows.  Section \ref{sec:background} outlines the architecture of our deliberation model and summarizes the techniques of MWER fine-tuning and shallow fusion.  Section \ref{sec:methods} describes the techniques we use to achieve the two goals given above.  Section \ref{sec:experiments} gives details on our dataset and model architecture.   Section \ref{sec:results} gives results and analysis, and we conclude in Section \ref{sec:conclusions}.
\section{Background}
\label{sec:background}

In this section, we summarize our baseline model, fine-tuning procedure, and method for language model integration.

\subsection{Deliberation Architecture}
Two-pass ASR models combine a pre-trained recurrent neural network transducer (RNN-T) \cite{Graves12} with a second decoder that rescores top-n hypotheses \cite{Chiu19}.  In a deliberation model, on the other hand, the second decoder has the option to attend to the RNN-T hypotheses instead of rescoring them, allowing all parts of the model to be trained together.

\begin{figure}[htb]
\centering
  \includegraphics[scale=0.33]{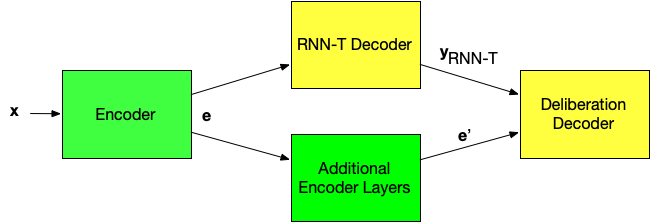}
  \caption{Deliberation Architecture, adapted from \cite{Hu20}.}
  \label{fig:Delib}
  \vspace{-0.1in}
\end{figure}

More specifically, a deliberation model's encoder consumes acoustic features $\textbf{x}$ and maps them onto encoder features $\textbf{e}$.  The RNN-T decoder attends to the encoder features, and an n-best list of hypothesis $\textbf{y}_\text{RNN-T}$ is extracted using a beam search.  A second encoder adapts $\textbf{e}$ into modified encoder features $\textbf{e}'$ to be consumed by the deliberation decoder.  The deliberation decoder attends to both features derived from $\textbf{y}_\text{RNN-T}$ and $\textbf{e}'$, and the final transcript is extracted with a second beam search.

\subsection{MWER Fine-tuning}
MWER training \cite{Shannon17} is a fine-tuning procedure designed to directly minimize the number of word errors instead of cross-entropy.  In MWER training, we seek to optimize the expected number of word errors over all possible hypotheses.  Since we cannot practically marginalize over all possible output sequences, we instead compute the expected word error rate from a sample of predictions:

\begin{equation}
L(\textbf{x}, \textbf{y}^*) = \sum_{\textbf{y} \in \textbf{B}} P(\textbf{y}|\textbf{x}) \hat{W}(\textbf{y}, \textbf{y}^*)
\end{equation}

where $\textbf{x}$ is the input acoustic features, $\textbf{y}^*$ is the ground truth, and $\hat{W}$ gives a normalized word error count.  Here, $\textbf{y}$ is a hypothesis from a beam $\textbf{B}$ that is sampled from the model using a beam search, and the posterior $P$ is normalized accordingly so that all probabilities sum to one.  It was demonstrated in \cite{Prabhavalkar17} that this method is effective for beam sizes as small as 4.

\subsection{Shallow Fusion}
In shallow fusion, a language model is incorporated into ASR decoding by interpolating the posteriors directly:

\begin{equation}
\label{eq:shallow-fusion}
\hat{\textbf{y}} = \argmax_{\textbf{y}} P_{\textbf{AM}}(\textbf{y} | \textbf{x}) + \alpha P_{\textbf{LM}}(\textbf{y}) + \beta \textbf{C} 
\end{equation}

where $\hat{\textbf{y}}$ is the selected hypothesis, $P_{\textbf{AM}}$ and $P_{\textbf{LM}}$ are posteriors in the acoustic and language models respectively, and $\alpha$ and $\beta$ are hyperparameters.  $\textbf{C}$ is a $\emph{coverage term}$ as in \cite{Chorowski16}, which seeks to discourage truncated transcripts by rewarding hypotheses that have been allocated weight above some threshold by the attention mechanism.

\section{Methods}
\label{sec:methods}

In this section, we describe our techniques for fusing an LM trained on our large text corpus into our deliberation model. 

\subsection{The Truncation Problem}
In \cite{Chorowski16}, the authors identified a failure mode for shallow fusion in which the fused model predicts a shortened transcript consisting of only the first few words that were spoken.  As we will see, this \say{truncation} problem turned out to be quite severe when incorporating our LMs, which were trained on a large text corpus.  

\subsubsection{Hyperparameters}

There are several hyperparameters proposed in the literature for the truncation problem.  We experimented with tuning the following:
\vspace*{-0.008in}
\begin{figure*}[!b]
  \begin{subfigure}[t]{0.5\columnwidth}
	  \includegraphics[width=1.0\columnwidth]{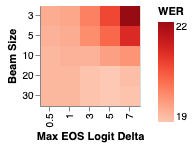}
	  \caption{\footnotesize Maps, LM Only}
  \end{subfigure}
  \begin{subfigure}[t]{0.5\columnwidth}
  	\includegraphics[width=1.0\columnwidth]{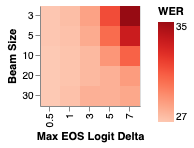}
	\caption{\footnotesize Search, LM Only}
  \end{subfigure}
  \begin{subfigure}[t]{0.5\columnwidth}  
	  \includegraphics[width=1.0\columnwidth]{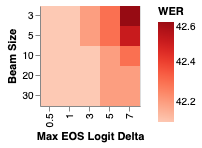}
	  \caption{\footnotesize Maps, Surprising Prons}
  \end{subfigure}
  \begin{subfigure}[t]{0.5\columnwidth}
	  \includegraphics[width=1.0\columnwidth]{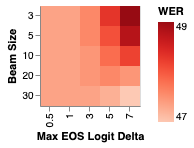}
	  \caption{\footnotesize Search, Surprising Prons}
  \end{subfigure}
  \begin{subfigure}[t]{0.53\columnwidth}
	  \includegraphics[width=1.0\columnwidth]{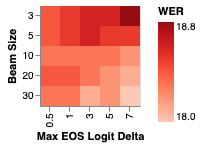}
	  \caption{\footnotesize Maps, LM Only}
  \end{subfigure}
  \begin{subfigure}[t]{0.53\columnwidth}
  	\includegraphics[width=1.0\columnwidth]{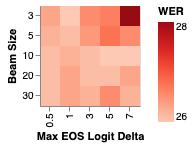}
	\caption{\footnotesize Search, LM Only}
  \end{subfigure}
  \begin{subfigure}[t]{0.52\columnwidth}  
	  \includegraphics[width=1.0\columnwidth]{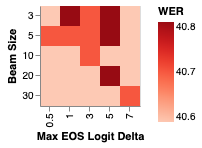}
	  \caption{\footnotesize Maps, Surprising Prons}
  \end{subfigure}
  \begin{subfigure}[t]{0.52\columnwidth}
	  \includegraphics[width=1.0\columnwidth]{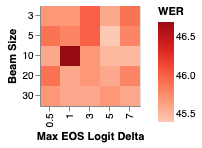}
	  \caption{\footnotesize Search, Surprising Prons}
  \end{subfigure}
  \caption{A sweep of maximum EOS log-probability delta and beam size on our test sets, before (a-d) and after (e-h)  MWER fine tuning.}
  \label{fig:sweep}
\end{figure*}

\begin{itemize}
	\item The coverage penalty $\textbf{C}$, as in equation \ref{eq:shallow-fusion} above.
	\item The beam size.  In principle, increasing the size of the beam leaves room for longer, less errorful hypotheses on the beam even when truncated hypotheses are present.
	\item The \emph{maximum EOS logprob delta}.  This hyperparameter is proposed in \cite{Chorowski16}.  When a hypothesis is ended with the EOS token during beam search, it must have a log probability no worse than that of the best hypothesis so far minus this value in order to be removed from the beam and marked as complete.
\end{itemize}

\subsubsection{MWER Fusion}
It was shown in \cite{McDermott18} that using a particular hyperparameter setting (blank scale, in that work) during training of a conventional ASR system can sometimes adapt the model to that setting in inference.  We attempt to reduce the difficulty of the hyperparameter tuning problem given above by showing that this can be done with beam search parameters in a deliberation model.

MWER fine tuning provides an opportunity to do this by running a beam search during training.  Since this beam search will include an LM in inference, we would like to perform shallow fusion during MWER fine tuning.  \cite{Weng19} develops a technique for fusion with RNN-T in which the LM's logit values are added to RNN-T's non-blank outputs, while leaving the logit for the blank output unchanged, and demonstrates small performance improvements.  Unlike \cite{Weng19}, this work seeks to use MWER training as a hyperparameter adaptation mechanism.  So, we instead fine tune the beam search of our second decoder, which will be the site of shallow fusion during inference.  Since this decoder does not emit a blank output, we can define a loss with direct logit interpolation:

\begin{equation}
L(\textbf{x}, \textbf{y}^*) = \sum_{\textbf{y} \in \textbf{B}_\text{LM}} (P_\textbf{AM}(\textbf{y} | \textbf{x}) + \alpha P_{\textbf{LM}} (\textbf{y}) + \beta \textbf{C}) \hat{W}(\textbf{y}, \textbf{y}^*)
\end{equation}

where $\textbf{B}_\text{LM}$ are hypotheses drawn using a beam search with shallow fusion.

\subsection{Taking Advantage of a Large Text Corpus}

\label{sec:language-modeling}
We implement the following pruning scheme to eliminate noisy data and reduce overfitting to extremely common sentences (e.g. \say{facebook}):

\begin{enumerate}
  \item For every sentence, each unigram is compared against a 1 million word vocabulary.  Any unigram not in this list is considered to be misspelled, and the sentence is discarded.
  \item If a sentence is duplicated $n$ times in the remaining examples, all but $\log(n)$ examples are discarded.
  \item The desired number of sentences are selected by random sampling. 
\end{enumerate}

Large text corpora have been exploited successfully for ASR in the past by sampling a training set that is relevant to a domain of interest \cite{Biadsy17}.  These results, however, used a maximum-entropy LM, which presents a convex optimization problem that scales naturally to large amounts of data, while we seek to optimize an non-convex RNN-LM.  Also, these results targeted geographical queries, which are plentiful in ASR training corpora, while we seek to improve performance on rare words.  Nevertheless, we seek to adapt this method to our problem by experimenting with an additional step between steps 2 and 3 above:

\begin{enumerate} [label=2*.]
  \item For every example, each unigram is compared to a list of word counts from the deliberation model's training data.  Any sentence not containing at least one \say{rare} word is discarded, where rareness is defined as occurring a number of times smaller than some threshold.
\end{enumerate}

All together, this scheme is designed to take advantage of the large size of our text corpus while still maintaining a manageable number of sentences for LM training.  We weigh the impact of this data reduction against that of increasing the LM's size.

\begin{table*}[!t]
	\centering
	\begin{tabular}{|c|c|c|c|c|}
		\toprule
		Experiment & Maps, LM Only & Search, LM Only & Maps, Surprising Prons & Search, Surprising Prons  \\
		\midrule
		\textbf{E1} & 18.8 & 27.3 & 42.1 & 47.6  \\
		\textbf{E2} & 18.8 & 27.1 & 42.1 & 47.4 \\
		\textbf{E2-4} & 18.7 & 26.9 & 41.9 & 47.1 \\
		\textbf{E2-6} & 18.6 & 27 & 41.8 & 47.2 \\
		\textbf{E2-8} & 18.6 & 26.9 & 41.9 & 47.1 \\
		\midrule
		\textbf{E3} & 18.7 & 26.5 & 41.2 & 46.6 \\
		\textbf{E4} & 18.7 & 26.6 & 41.4 & 46.7 \\
		\bottomrule
	\end{tabular}
    \vspace{0.1in}
    \caption{WER Results of Expanded Models}
	\label{tab:results}
	\vspace{-0.8cm}
\end{table*}

\section{Experiments}
\label{sec:experiments}

In this section, we describe the parameters of our experiments.  We also describe our methods for measuring the success of our LM integration and evaluating performance on the tail.

\subsection{Deliberation Model Training}
Our deliberation model is similar to that presented in \cite{Hu20}.  We use 128-dimensional log-Mel audio features with a 32ms window and 10ms shift.  The RNN-T component of the deliberation model contains eight LSTM layers in its encoder, each with 2,048 units and a 640-dimensional projection.  The joint network contains 640 units, followed by a final softmax layer.  Hypotheses from RNN-T are passed to a two-layer bidirectional LSTM which projects them into a 320-dimensional space.  Our second decoder attends to both these features and RNN-T encoder output and emits context vectors which are passed to a final 2-layer LSTM.

Our training set is described in \cite{Narayanan18}.   Transcripts are lower-cased and processed with a 4k word piece model.
\subsection{Language Model Training}
Our language models are similar to those in \cite{Toshniwal18}.  The models consist of LSTM layers with 512 nodes each, with a projection layer of 256 nodes.  Our baseline model has two hidden layers.

The models are trained on a sample of anonymized production traffic to Google applications.  We divide this data into domains that describe the origin of the queries.  All examples are stripped of metadata, so that only the query text is visible to the model.  Our training set is selected from this data using the pruning procedure outlined in section \ref{sec:language-modeling}.  The total size of our data before pruning is about 230 billion examples.  Vocabulary pruning (Step 1) reduces that size to about 218 billion, and $\log{n}$ pruning (Step 2) further reduces the size to about 25 billion examples.

For our baseline models, we omit rare word filtering (Step 2*) from our pruning procedure and sample down to a final size of 4.5 billion examples (about 50 billion words).  When we include rare word filtering, we obtain a dataset of about 1 billion total examples (about 11 billion words), and omit Step 3.

\subsection{Evaluation Sets}
We would like to create evaluation sets that measure the degree to which our LM has been integrated into our model, and to determine performance on the tail.  We create separate test sets for these two purposes.  We split our test sets into those focused on geographical queries (Maps) and general queries (Search). The test sets are created by looking for utterances in the text data that have a very different perplexity distribution compared to the audio-text pair training data.

To measure LM integration, we build test sets consisting of words that are common in the LM training data but rare in the AM training data.  To this end, we compute unigram statistics for both corpora and construct a list of unigrams that occur at most five times in the AM data (about three quarters of all words) and at least 150 times in the LM data (about 99\% of all words).

To measure tail performance, we target words that have pronunciations that are surprising given the spelling.  Unusual pronunciations have been shown to be difficult for ASR systems \cite{Peyser20, Beaufays03, Laurent10}.  To select examples with surprising utterances, we manually assemble a map from grapheme sequences to corresponding phoneme sequences.  Our mapping consists of 487 correspondences.  For a given example, we process each unigram grapheme-by-grapheme, using the map to assemble a list of possible corresponding phoneme sequences. If none of the predicted pronunciations match the true pronunciation of the unigram, we consider the unigram to have surprising pronunciation.

For each test set, we select 10000 examples and  synthesize audio for each transcript with a TTS system as in \cite{Gonzalvo16}. 

\begin{table}[!b]
	\vspace{-0.3cm}
	\centering
	\begin{subtable}[t]{0.4\textwidth}
		\centering
		\begin{tabular}{|c|c|c|}
			\toprule
			              & WER & Truncation WER \\
			\midrule
			Baseline & 19.9 & 2.0 \\
			Fusion   &  21.0 & 3.6 \\
			Fusion w/ BS Params & 18.7 & 2.2 \\
			\bottomrule
		\end{tabular}
		\caption{Maps, LM Only}
	\end{subtable}
	\hfill
	\begin{subtable}[t]{0.4\textwidth}
		\centering
		\begin{tabular}{|c|c|c|}
			\toprule
			              & WER & Truncation WER \\
			\midrule
			Baseline & 28.4 & 6.9 \\
			Fusion   &  31.5 & 7.6 \\
			Fusion w/ BS Params & 27.1 & 7.1 \\
			\bottomrule
		\end{tabular}
		\caption{Search, LM Only}
	\end{subtable}	
	\caption{Impact of the Truncation Problem}
	\label{tab:truncation}
\end{table}

\section{Results}
\label{sec:results}

This section presents our experimental results and discussion.

\subsection{Truncation}
We find that optimizing only the LM interpolation weight $\alpha$ and coverage penalty weight $\beta$ does not yield improvement over our baseline model.  Improvement was only shown after tuning beam size and maximum EOS logprob delta, which relate directly to the beam search.  To understand the problem, we compare our models' word error rate to \say{truncation word error rate}, which is the word error rate on examples for which the prediction has at most half as many unigrams as the reference.  Table \ref{tab:truncation} compares WER and Truncation WER for our baseline deliberation model to a fusion model with $\alpha=0.1$ and $\beta=0.06$ and to a second fusion model in which the beam size is set to $20$ and maximum EOS logprob delta is set to $0.05$.  This data suggests that truncation errors are largely responsible for the degradation in WER in the initial fusion model, and that tuning the beam search parameters recovers those losses.

Figure \ref{fig:sweep} (a-d) shows the results of a sweep of the two beam search parameters: beam size and maximum EOS logprob delta.  Interestingly, while we find that increasing beam size yields improvements, for a sufficiently small value of maximum EOS logprob delta the beam size does not make a difference.  Nevertheless, it is clear that WER results are highly dependent on correct setting of these hyperparameters.

We find that MWER fine-tuning dramatically diminishes the importance of beam search parameters in evaluation.  Figure \ref{fig:sweep} (e-h) shows the results of training 25 MWER models, using the same combinations of maximum EOS logprob delta and beam size from above during the MWER beam search and then evaluating using shallow fusion with those same parameters.  We find a significantly smaller range of WER than before MWER fine tuning.  This suggests that MWER fine tuning serves to adapt a model to some choice of beam search parameters by using those parameters during training. This could make MWER useful as a tool to alleviate the difficulty of hyperparameter tuning in shallow fusion.

\subsection{Language Model Size}
We compare the importance of model size to data selection criterion in LM training.  Table \ref{tab:results} gives results for shallow fusion with four progressively larger LMs including our baseline (\textbf{E1}), an expanded model in which the LM's projection layer is removed effectively doubling the parameter count (\textbf{E2}), and that same model with 4, 6, and 8 hidden layers (\textbf{E2-4, E2-6, E2-8}).  Table \ref{tab:results} also gives results for the 4-layer variant in which Step 2* of the pruning procedure described in Section \ref{sec:language-modeling} is applied to the training data (\textbf{E3}), and finally where the training set is further sampled down to about 50 million examples (\textbf{E4}).

We see that filtering rare words shows significantly larger gains than increasing model capacity. This suggests that it is easier to take advantage of a large text corpus by selecting a subset of relevant examples than it is to model the entire distribution.  Interestingly, this benefit is strongest when we only prune to 1 billion examples, and weakens when we further prune down to 50 million.  This further suggests that an RNN-LM used in fusion is capable of benefiting from a very large text corpus.

\section{Conclusions}
\label{sec:conclusions}
In this paper, we've explored shallow fusion using a very large text-only corpus.  We've quantified and explored solutions to the truncated utterances problem and demonstrated that MWER fine tuning almost eliminates the need for hyperparameter tuning.  Finally we showed how a pruning strategy can beat out large models in taking advantage of large amounts of text data.

\bibliographystyle{IEEEtran}

\bibliography{Main}

\begin{thebibliography}{10}
\providecommand{\url}[1]{#1}
\csname url@samestyle\endcsname
\providecommand{\newblock}{\relax}
\providecommand{\bibinfo}[2]{#2}
\providecommand{\BIBentrySTDinterwordspacing}{\spaceskip=0pt\relax}
\providecommand{\BIBentryALTinterwordstretchfactor}{4}
\providecommand{\BIBentryALTinterwordspacing}{\spaceskip=\fontdimen2\font plus
\BIBentryALTinterwordstretchfactor\fontdimen3\font minus
  \fontdimen4\font\relax}
\providecommand{\BIBforeignlanguage}[2]{{%
\expandafter\ifx\csname l@#1\endcsname\relax
\typeout{** WARNING: IEEEtran.bst: No hyphenation pattern has been}%
\typeout{** loaded for the language `#1'. Using the pattern for}%
\typeout{** the default language instead.}%
\else
\language=\csname l@#1\endcsname
\fi
#2}}
\providecommand{\BIBdecl}{\relax}
\BIBdecl

\bibitem{Sak13}
H.~{Sak}, F.~{Beaufays}, K.~{Nakajima}, and C.~{Allauzen}, ``Language model
  verbalization for automatic speech recognition,'' in \emph{IEEE ICASSP},
  2013, pp. 8262--8266.

\bibitem{Huang10}
S.~{Huang} and S.~{Renals}, ``Hierarchical bayesian language models for
  conversational speech recognition,'' \emph{IEEE Transactions on Audio,
  Speech, and Language Processing}, vol.~18, no.~8, pp. 1941--1954, 2010.

\bibitem{Wang19}
D.~Wang, X.~Wang, and S.~Lv, ``An overview of end-to-end automatic speech
  recognition,'' \emph{Symmetry}, vol.~11, p. 1018, 08 2019.

\bibitem{He18}
Y.~{He}, T.~N. {Sainath}, R.~{Prabhavalkar}, I.~{McGraw}, R.~{Alvarez},
  D.~{Zhao}, D.~{Rybach}, A.~{Kannan}, Y.~{Wu}, R.~{Pang}, Q.~{Liang},
  D.~{Bhatia}, Y.~{Shangguan}, B.~{Li}, G.~{Pundak}, K.~C. {Sim}, T.~{Bagby},
  S.~{Chang}, K.~{Rao}, and A.~{Gruenstein}, ``Streaming end-to-end speech
  recognition for mobile devices,'' in \emph{IEEE ICASSP}, 2019, pp.
  6381--6385.

\bibitem{Caglar15}
\BIBentryALTinterwordspacing
{\c{C}}.~G{\"{u}}l{\c{c}}ehre, O.~Firat, K.~Xu, K.~Cho, L.~Barrault, H.~Lin,
  F.~Bougares, H.~Schwenk, and Y.~Bengio, ``On using monolingual corpora in
  neural machine translation,'' \emph{CoRR}, vol. abs/1503.03535, 2015.
  [Online]. Available: \url{http://arxiv.org/abs/1503.03535}
\BIBentrySTDinterwordspacing

\bibitem{Sriram17}
A.~Sriram, H.~Jun, S.~Satheesh, and A.~Coates, ``Cold fusion: Training seq2seq
  models together with language models,'' in \emph{INTERSPEECH}, 2018.

\bibitem{Toshniwal18}
S.~Toshniwal, A.~Kannan, C.-C. Chiu, Y.~Wu, T.~Sainath, and K.~Livescu, ``A
  comparison of techniques for language model integration in encoder-decoder
  speech recognition,'' \emph{IEEE SLT}, 2018.

\bibitem{McDermott19}
E.~{McDermott}, H.~{Sak}, and E.~{Variani}, ``A density ratio approach to
  language model fusion in end-to-end automatic speech recognition,'' in
  \emph{2019 IEEE Automatic Speech Recognition and Understanding Workshop
  (ASRU)}, 2019, pp. 434--441.

\bibitem{Chorowski16}
J.~Chorowski and N.~Jaitly, ``Towards better decoding and language model
  integration in sequence to sequence models,'' \emph{INTERSPEECH}, 2016.

\bibitem{Battenberg17}
E.~Battenberg, J.~Chen, R.~Child, A.~Coates, Y.~Li, H.~Liu, S.~Satheesh,
  A.~Sriram, and Z.~Zhu, ``Exploring neural transducers for end-to-end speech
  recognition,'' \emph{2017 IEEE Automatic Speech Recognition and Understanding
  Workshop (ASRU)}, pp. 206--213, 2017.

\bibitem{Povey11}
D.~Povey, A.~Ghoshal, G.~Boulianne, L.~Burget, O.~Glembek, N.~Goel,
  M.~Hannemann, P.~Motlícek, Y.~Qian, P.~Schwarz, J.~Silovský, G.~Stemmer,
  and K.~Vesel, ``The kaldi speech recognition toolkit,'' \emph{IEEE 2011
  Workshop on Automatic Speech Recognition and Understanding}, 01 2011.

\bibitem{Rafal16}
\BIBentryALTinterwordspacing
R.~J{\'{o}}zefowicz, O.~Vinyals, M.~Schuster, N.~Shazeer, and Y.~Wu,
  ``Exploring the limits of language modeling,'' \emph{CoRR}, vol.
  abs/1602.02410, 2016. [Online]. Available:
  \url{http://arxiv.org/abs/1602.02410}
\BIBentrySTDinterwordspacing

\bibitem{Chelba13}
C.~Chelba, T.~Mikolov, M.~Schuster, Q.~Ge, T.~Brants, P.~Koehn, and
  T.~Robinson, ``One billion word benchmark for measuring progress in
  statistical language modeling,'' \emph{INTERSPEECH}, vol. abs/1312.3005,
  2014.

\bibitem{Radford19}
\BIBentryALTinterwordspacing
A.~Radford, J.~Wu, R.~Child, D.~Luan, D.~Amodei, and I.~Sutskever, ``Language
  models are unsupervised multitask learners,'' 2019. [Online]. Available:
  \url{https://openai.com/blog/better-language-models/}
\BIBentrySTDinterwordspacing

\bibitem{Kannan18}
A.~Kannan, Y.~Wu, P.~Nguyen, T.~Sainath, Z.~Chen, and R.~Prabhavalkar, ``An
  analysis of incorporating an external language model into a
  sequence-to-sequence model,'' in \emph{IEEE ICASSP}, 04 2018, pp. 1--5828.

\bibitem{Hu20}
K.~Hu, T.~Sainath, R.~Pang, and R.~Prabhavalkar, ``Deliberation model based
  two-pass end-to-end speech recognition,'' in \emph{IEEE ICASSP}, 2020.

\bibitem{Graves12}
A.~Graves, ``Sequence transduction with recurrent neural networks,'' in
  \emph{International Conference of Machine Learning (ICML) Workshop on
  Representation Learning}, vol. abs/1211.3711, 2012.

\bibitem{Chiu19}
C.-C. Chiu, D.~Rybach, I.~McGraw, M.~Visontai, Q.~Liang, R.~Prabhavalkar,
  R.~Pang, T.~Sainath, T.~Strohman, W.~Li, Y.~R. He, and Y.~Wu, ``Two-pass
  end-to-end speech recognition,'' in \emph{Interspeech}, 2019.

\bibitem{Shannon17}
M.~Shannon, ``Optimizing expected word error rate via sampling for speech
  recognition,'' in \emph{INTERSPEECH}, 2017.

\bibitem{Prabhavalkar17}
R.~Prabhavalkar, T.~N. Sainath, Y.~Wu, P.~Nguyen, Z.~Chen, C.-C. Chiu, and
  A.~Kannan, ``Minimum word error rate training for attention-based
  sequence-to-sequence models,'' in \emph{IEEE ICASSP}, 2018, pp. 4839--4843.

\bibitem{McDermott18}
\BIBentryALTinterwordspacing
E.~McDermott, ``A deep generative acoustic model for compositional automatic
  speech recognition,'' in \emph{Proceedings of Neural Information Processing
  Systems (NeurIPS) Workshop: Interpretability and Robustness in Audio, Speech,
  and Language}, 2018. [Online]. Available:
  \url{https://openreview.net/pdf?id=S1fbqB0noQ}
\BIBentrySTDinterwordspacing

\bibitem{Weng19}
C.~Weng, C.~Yu, J.~Cui, C.~Zhang, and D.~Yu, ``Minimum bayes risk training of
  rnn-transducer for end-to-end speech recognition,'' \emph{ArXiv}, vol.
  abs/1911.12487, 2019.

\bibitem{Biadsy17}
F.~Biadsy, M.~Ghodsi, and D.~Caseiro, ``Effectively building tera scale maxent
  language models incorporating non-linguistic signals,'' in
  \emph{INTERSPEECH}, 2017.

\bibitem{Narayanan18}
A.~Narayanan, A.~Misra, K.~C. Sim, G.~Pundak, A.~Tripathi, M.~Elfeky,
  P.~Haghani, T.~Strohman, and M.~Bacchiani, ``Toward domain-invariant speech
  recognition via large scale training,'' in \emph{IEEE Spoken Language
  Technology Workshop (SLT)}, 2018.

\bibitem{Peyser20}
C.~{Peyser}, T.~N. {Sainath}, and G.~{Pundak}, ``Improving proper noun
  recognition in end-to-end asr by customization of the mwer loss criterion,''
  in \emph{IEEE ICASSP}, 2020, pp. 7789--7793.

\bibitem{Beaufays03}
F.~{Beaufays}, A.~{Sankar}, S.~{Williams}, and M.~{Weintraub}, ``Learning name
  pronunciations in automatic speech recognition systems,'' in
  \emph{Proceedings. 15th IEEE International Conference on Tools with
  Artificial Intelligence}, Nov 2003, pp. 233--240.

\bibitem{Laurent10}
A.~Laurent, S.~Meignier, T.~Merlin, and P.~Deleglise, ``Acoustics-based
  phonetic transcription method for proper nouns,'' in \emph{Proceedings of the
  11th Annual Conference of the International Speech Communication Association,
  INTERSPEECH 2010}, 01 2010, pp. 2286--2289.

\bibitem{Gonzalvo16}
\BIBentryALTinterwordspacing
X.~Gonzalvo, S.~Tazari, C.~Chan, M.~Becker, A.~Gutkin, and H.~Sil{\'{e}}n,
  ``Recent advances in google real-time hmm-driven unit selection
  synthesizer,'' in \emph{Interspeech 2016, 17th Annual Conference of the
  International Speech Communication Association, San Francisco, CA, USA,
  September 8-12, 2016}, N.~Morgan, Ed.\hskip 1em plus 0.5em minus 0.4em\relax
  {ISCA}, 2016, pp. 2238--2242. [Online]. Available:
  \url{https://doi.org/10.21437/Interspeech.2016-264}
\BIBentrySTDinterwordspacing

\end{thebibliography}

\end{document}